\documentclass[10pt]{article}
\usepackage[]{times,psfig,emulateapj}

\def\gapp{\ifmmode\stackrel{>}{_{\sim}}\else$\stackrel{>}{_{\sim}}$\fi}
\def\lapp{\ifmmode\stackrel{<}{_{\sim}}\else$\stackrel{<}{_{\sim}}$\fi}
\def\degr{\hbox{$^\circ$}}

\lefthead{Johnston et al.}
\righthead{High time-resolution observations of the Vela pulsar}

\begin{document}

\title{High time-resolution observations of the Vela pulsar} 

\author{Simon Johnston\altaffilmark{1},
Willem van Straten\altaffilmark{2}, Michael Kramer\altaffilmark{3},
and Matthew Bailes\altaffilmark{2} }

\altaffiltext{1}{Research Centre for Theoretical Astrophysics, University of
Sydney, NSW 2006, Australia}

\altaffiltext{2}{Swinburne Centre for Astrophysics and Supercomputing,
Swinburne University of Technology, Hawthorn, Vic 3122, Australia}

\altaffiltext{3}{University of Manchester, Jodrell Bank Observatory,
Macclesfield, Cheshire SK11 9DL, UK}
%
%
%
%
%
\authoraddr{Address correspondence regarding this manuscript to: 
		Simon Johnston
		School of Physics
		University of Sydney
	        NSW 2006,
		Australia}
\begin{abstract}
We present high time resolution observations of single pulses
from the Vela pulsar (PSR B0833--45) made with a baseband recording system
at observing frequencies of 660 and 1413 MHz.  We have discovered 
two startling features in the 1413 MHz single pulse data. The first is the
presence of giant micro-pulses which are confined to the leading edge
of the pulse profile. One of these pulses has a peak flux density in excess
of 2500 Jy, more than 40 times the integrated pulse peak.
The second new result is the presence of a
large amplitude gaussian component on the trailing edge of the pulse
profile. This component can exceed the main pulse in intensity but is
switched on only relatively rarely.
Fluctutation spectra reveal
a possible periodicity in this feature of 140 pulse periods.
Unlike the rest of the profile, this component
has low net polarization and emits predominantly in the orthogonal mode.
This feature appears to be unique to the Vela pulsar.
We have also detected microstructure in the Vela pulsar for the first time.
These same features are present in the 660 MHz data.
We suggest that the full width of the Vela pulse profile might be as large
as 10 ms but that the conal edges emit only rarely.
\end{abstract}


\section{Introduction}
The Vela pulsar is one of the closest and brightest radio pulsars
known.  Polarization observations made shortly after its discovery
showed that it is highly linearly polarized and that the position
angle of the radiation followed an S-shape curve as a function of
pulse longitude (Radhakrishnan \& Cooke 1969). This showed that the
radio emission was related to the geometry of the magnetic field lines
near the pole. The rotating vector model (RVM) of Radhakrishnan \&
Cooke has been extensively used in a large number of pulsars to obtain
the spin axis - magnetic axis and the viewing angle - magnetic axis
geometries. In the Vela pulsar, the angle between the spin and magnetic axes,
$\alpha$ is estimated to lie between 60\degr\ and 90\degr, and the
impact angle, $\beta$ is $\sim -6\degr$ (Krishnamohan \& Downs 1983
(hereafter KD83); Lyne \& Manchester 1988; Rankin 1993).

In the only detailed single pulse study of
Vela to have appeared in the (recent) literature,
KD83 observed 87000 pulses at 2.3 GHz with
a time resolution including dispersion smearing of 750$\mu$s.
Their main conclusion was that the pulse profile consists of four
different components which were emitted at different heights in the
magnetosphere. They also found that the pulse shape depended on the
intensity with the strongest pulses arriving early with respect to the
average profile.

In some pulsars, the S-shaped sweep of polarization is
broken by one or more jumps of $90\degr$. These are known as
`orthogonal' jumps (Backer \& Rankin 1980). It is generally believed
that emission from two orthogonally polarized modes are superposed (McKinnon
\& Stinebring 1998). At a given pulse longitude the dominant mode then
determines the position angle. In the Vela pulsar, KD83 found that only
one mode was present at all pulse longitudes for all 87000 pulses.

Sub-pulse structure is ubiquitous in pulsars on a number of different
timescales. So-called microstructure is often seen and can have both
a typical width (generally a few tens of $\mu$s) and be quasi-periodic 
in a given pulsar (e.g. Lange et al. 1998). Surprisingly, for the Vela
pulsar no microstructure analysis has ever been published.

The phenomenon of giant pulses has only been detected in two pulsars,
the Crab pulsar (e.g. Lundgren et al. 1995) and PSR B1937+21 (Cognard
et al. 1996).  The working definition of giant pulses is a flux
density in a single pulse which is more than 10 times the mean flux
density. Both the Crab and PSR B1937+21 show occasional pulses in
excess of 1000 times the mean pulsed flux. In the vast majority of
pulsars, very few if any single pulse fluxes exceed 10 times the mean
flux density.  The Crab is one of the youngest known pulsars whereas PSR
B1937+21 is a millisecond pulsar with very rapid rotation
rate.  As Cognard et al. (1996) point out, the common feature these
pulsars share is the highest (estimated) magnetic field at their light
cylinder, although whether this is related to the physics of the giant
pulses is unclear.  In this parameter, Vela ranks 22nd in the current
catalogue of 965 pulsars.

\section{Observations and Data Reduction}

As part of a project investigating single pulses from a large sample
of pulsars, we observed the Vela pulsar at two different observing
frequencies between 2000 March 14 and 17 using the 64-m Parkes
radiotelescope. The center frequencies of the observations were 660
and 1413 MHz; at these frequencies the system equivalent flux
density is 120 and 26 Jy respectively.  Each receiver consists of two
orthogonal feeds sensitive to linear polarization. The signals are
down-converted and amplified before being passed into the backend. The
backend, CPSR, is an enhanced version of the Caltech Baseband Recorder
(Jenet et al. 1997).  It consists of an analogue dual-channel
down-converter and digitizer card which yields 2-bit quadrature 
samples at 20 MHz.  The data stream is
written to DLT for subsequent off-line processing allowing all 4 Stokes
parameters to be computed. At both
frequencies, 30 minutes of data or $\sim$20000 pulses were
recorded. Before each observation, a 90-s observation of a pulsed
signal, directly injected into the receiver at a $45\degr$ angle to the feed,
is made. This enables instrumental polarization to be
corrected. Observations of the flux calibrator Hydra A were made at
each of the observing frequencies; this allows absolute fluxes to be
obtained.

The data were processed off-line using a workstation cluster at the
Swinburne Supercomputer Centre. Data reduction involves
coherent de-dispersion (Hankins \& Rickett 1975)
and includes quantization error corrections as described by
Jenet \& Anderson (1998). The data are folded at the apparent 
topocentric period of the pulsar and the full Stokes profiles for each
pulse are written to disk. Flux calibration and instrumental calibration
are then carried out using information contained in the observation
of the pulsed (calibration) signal. The data in each frequency channel 
is corrected for the rotation measure of the pulsar and all the 
frequency channels are then summed to produce the final profile.
At 1413 MHz there are 2048 time-bins per pulse period for an effective 
time resolution of 44 $\mu$s, comparable to the
scatter broadening at this frequency. At the lower frequency, the
time resolution is 88 $\mu$s, but the
scatter broadening dominates and the effective time resolution is
$\sim$0.5 ms at 660 MHz.

\section{Results}
\subsection{Microstructure}

As far as we are aware, the literature has not pointed out the
existence of microstructure in the Vela pulsar, generally because the
time resolution of the observations has been too coarse.
Figure 1 shows `typical' single pulses from Vela at 1413 MHz with a time
resolution of 44$\mu$s. Microstructure is clearly seen. This microstructure
is ubiquitous in virtually every pulse at 1413 MHz. We conclude that
the smooth pulse profiles shown by KD83 are an artifact of 
their coarse time resolution.

At lower observing frequencies, pulse scatter broadening smooths
over the microstructure features, although they are still visible in
the 660 MHz data. A complete investigation of the microstructure
in the Vela pulsar will be presented elsewhere (Kramer et al. 2001).

\subsection{Pulse Intensity}

Figure 2 shows a histogram of the mean flux density for the 20085
pulses recorded at 1413 MHz. More than 95\% of the pulses are within a
factor of 2 of the mean flux density, 99.5\% are within a factor of 3
and there are no pulses greater than 10 times the mean flux density.
This distribution is typical of many, perhaps most pulsars.  There are
no giant pulses in Vela in the same sense as those in the Crab pulsar
and PSR B1937+21.  It is well known that Vela does not show any nulls
(e.g. KD83; Biggs 1986), and indeed we also see no evidence of nulling in the
$\sim$40,000 pulses collected at the two observing frequencies. The
weakest single pulses have continuum flux densities of 1.3 and 0.3 Jy at
660 and 1413 MHz respectively.

We computed the modulation index, $\sigma_i$/$m_i$ where $\sigma_i$ is
the rms and $m_i$ the mean intensity in the $i$th bin, for each bin in
the pulse profile.  As also shown by KD83, the modulation index is low
in the centre of the pulse profile and increases towards the wings. To
show this effect in a more striking way, we computed the quantity
$R_i$=(MAX$_i$--$m_i$)/$\sigma_i$ where MAX$_i$ is the maximum
intensity in the $i$th bin. $R_i$ as a function of pulse phase is
shown in Figure 3. It can be seen that the value of $R$ is consistent
with Gaussian statistics in the centre of the pulse but that both the
leading and trailing edges of the pulse have individual pulses which
are extremely high amplitude with respect to the mean intensity. A
very similar result is seen at 660 MHz.  The three main features in
this plot will be discussed in more detail below.

\subsection{Orthogonal Modes}

The sweep of the position angle across the pulse in the Vela pulsar
shows the characteristic S shape without any of the `orthogonal' jumps
seen in a large number of other pulsars. The swing of the position
angle is well fitted by the RVM and shows that the line-of-sight cuts
very close to the magnetic axis, i.e.~$\alpha\sim55\degr$ and
$\beta\sim-6\degr$. To investigate whether
any individual pulse shows emission in the orthogonal mode we compared
the position angle of each bin in each pulse to that of the position
angle in the integrated profile at the same pulse phase.  Of the
40,000 pulses recorded at 660 and 1413 MHz, {\em none} shows evidence
for orthogonal mode emission in the region of the profile's peak,
consistent with the results of KD83 at 2295 MHz. 

However, investigating the feature centered near phase +4 ms in Figure 3,
we discovered that it is a result of only 21 pulses at 1413 MHz.
These pulses not only exhibit integrated flux densities more than
10 times that of the integrated profile at this longitude, but they
also have a consistently small fractional polarization.  Similarly, at
660 MHz there are 23 pulses with the same characteristics.  Of these,
all had low net polarization but, where polarization was present it
was orthogonal to the integrated polarization at this pulse phase.
Figure 4 shows the profile resulting from the summation of the 21
pulses taken at 1413 MHz. The large intensity of this `bump' feature
compared to the integrated profile and the orthogonal mode emission
can clearly be seen. We also note that the circular polarization is
negative under the bump emission and this is of the opposite sign to
that in the integrated profile, as is commonly seen in orthogonal mode
emission.  Introducing a jump of $90\degr$ in the position angle of
the bump region results in the identical results for $\alpha$ and
$\beta$ when fitting the RVM.

The midpoint of the `bump' emission occurs 3.86 ms after the main pulse
peak at 1413 MHz and 4.37 ms afterwards at 660 MHz, a significant difference.
Thus the profile is wider at 660 MHz than at 1413 MHz. The
`bump' component is also stronger relative to the main pulse at 1413
MHz than at 660 MHz by nearly a factor of 2.
This frequency dependence is in the direction
expected in the radius-to-frequency mapping paradigm.  Such a high
intensity, relatively rarely-appearing component has not been detected
in any other pulsar although the interpulse in PSR B0950+08 may show
similar properties.

As stated above, at 1413 MHz there are 21 occurrences of the bump
component in the $\sim$20000 pulses. A look at the pulse numbers, however,
reveals a far from random sequence. There are 7 cases in which the
differences in pulse number between appearances of the bump
component is less than 300.  Of these, 5 pairs have
intervals either in the range 130-140 pulses or 260-280 pulses and
other related harmonics are also present. We computed fluctuation
spectra from the single pulses, producing results which are somewhat
different to those of KD83, probably due to our much better time
resolution. Including 15360 pulses in our analysis, we indeed find
periodicities of $\sim 0.007$ and $\sim0.025$ cycles/period,
i.e.~every $\sim142$ and $\sim40$ pulses, in the `bump' region. These
periodicities are significant at the $\ga$6$\sigma$ level.
We also detect an even lower frequency feature in the main peak region, which
could be associated with a `drift' in micropulses, which we sometimes observe. 
A longer dataset is required to determine the significance of these
features; observations at 1400 and 2400 MHz will be carried out in the
near future.

\subsection{Giant Micro-pulses}

At 1413 MHz there are 14 pulses for which $R>25$ and which do not
occur in the bump region. The individual pulses are shown
superimposed in Figure 5.  Of these 14, the two which occur earliest
in pulse phase are both centered at phase $-2.2$ ms (relative to the
peak of the integrated profile) and are very narrow with a full-width
at half-maximum of only $\sim$50$\mu$s. The other 12 pulses have peak
phases between $-1.4$ and $-1.8$ ms and have typical half-widths of
$\sim$400$\mu$s. The brightest of these pulses has a peak flux 
density in
excess of 2500 Jy, more than 40 times the peak flux density
in the integrated
profile and more than 130 times the mean flux density
at this pulse phase.  At
660 MHz the results are very similar. One high amplitude pulse is seen
at early phases and 19 at phases just prior to the main pulse.  The
peak amplitudes are again in excess of 2000 Jy, this is only a factor
of $\sim$10 higher than the peak of the main pulse at this frequency.

We call all these features giant micro-pulses : they are not true giant
pulses because their mean flux density does not exceed 10 times
the mean flux density, but their peak flux densities are very large in
absolute terms and they appear far more often than would be expected
from any `normal' flux probability distribution.  The polarization of
these giant micro-pulses is extremely high with a linear fraction
close to 100\%. Furthermore, the position angle of the polarization is
exactly that expected by extrapolating the main position angle swing
back to these longitudes. Finally, 13 of the 14 pulses have small
negative circular polarization and one has positive circular
polarization.

Although KD83 also noted that strong intensity pulses arrived at early
phases, they make no explicit mention of single giant pulses.
However, they seem to have employed a time-averaging technique which
may have had the effect of broadening the giant pulses (because they
do not all arrive at the same pulse phase) and reducing their
amplitude. Our secondary feature more than 2 ms earlier
than the main pulse peak was not covered in their `on-pulse' bins.

\section{Discussion}

If we assume that the line-of-sight crosses the magnetic axis at the
peak of the profile, we can compute the emission height at 660 and
1413 MHz by measuring the delay of the steepest swing of the position
angle relative to the profile peak (Blaskiewicz et al. 1991; Hibschman
\& Arons 2000).  Using the result of our RVM fits, the 
delay is 1.3 ms at 660 MHz and 1.04 ms at 1413 MHz
leading to emission heights of 98 and 78 km respectively, although
the 660 MHz value has a significant error bar due to the scatter
broadening.  These emission heights
are typical in pulsars, and the frequency scaling is also within the
range seen in other pulsars (e.g.~Kramer et al.~1994).  In turn, by
assuming dipolar field lines, the emission heights can be used to
derive the half-opening angle of the cone, $\rho=3\sqrt{\pi \; r_{em}
/(2\; c \;P)}$, respectively 13$\degr$
and 11.6$\degr$ at the two frequencies.

Knowledge of $\rho$ and the viewing geometry can be used to derive the
width of the pulse profile. (Normally this process is used in reverse,
with the known pulse width used to derive $\rho$).  For $\alpha \sim
55\degr$ and $\beta \sim -6\degr$, the pulse full-width should be
$\sim30\degr$ or $\sim7.4$ ms. Even when considering that this width
is to be measured at a very low intensity level, it seems much larger
than observed in the integrated profile.  However, the emission from
the `bump' component now extends the pulse profile towards the trailing
edge, matching the expected width quite nicely. Luo \& Melrose (2001)
have recently proposed that the trailing edges of wide cones have their
emission suppressed due to cyclotron absorption within the pulsar
magnetosphere. Such a mechanism may explain why these `bump' pulses
are seen so rarely.

The presented calculation assumes that the magnetic axis coincides with
the pulse peak, tempting the speculation that we are `missing` the
leading edge of the cone. Perhaps extending to a phase of $-5$ ms, the
cone may not emit at all or only very rarely. We note that the giant
micro-pulses do increase the pulse width on the leading edge but only
by 1 ms or so. 

One could also assume that, with the extra bump component, we
are seeing the full pulse width. In this case, the magnetic pole
crossing would be close to phase $+2$ ms. However, this reasoning
does not fit well with a number of the observations. First, the 
steepest swing of the position angle sweep occurs {\it prior} to this
point. Also, the frequency dependence of the pulse width and the
spectral index of the various features show that the `main' pulse
is more likely to be the core component and the `bump' more likely
to be conal (e.g.~Kramer et al.~1994).

Considering the apparent two groups of giant micro-pulses, it is
unclear whether we are dealing with two separate phenomenon or whether
the small number statistics are preventing us from seeing these
features at all pulse phases near $-2$ ms.  However, the giant
micro-pulses at the earliest phases are of much shorter duration than
those occuring later, and the same pattern is seen at 660 MHz as at
1413 MHz. What is certain is that they all lead the main pulse; there
are no such features anywhere after the main pulse peak.

These giant micro-pulses seem to have more in common with the giant
pulses in PSR B1937+21 than with the Crab giant pulses. In the former,
the giant pulses all occur at the same pulse phase, whereas in the Crab
the giant pulses can arrive anywhere within the pulse window.
In Vela, the giant micro-pulses all arrive within a narrow phase range
and lead the main pulse (unlike in PSR B1937+21 where the giant pulses
lag the main pulse).

\acknowledgments We thank R. Edwards for help with the observations.
The Australia Telescope is funded by the Commonwealth of Australia
for operation as a National Facility managed by the CSIRO.

\clearpage
\newpage
\begin{figure}
\caption{Two individual pulses at 1413 MHz showing microstructure. 
Zero phase refers to the peak of the integrated profile.}
\psfig{figure=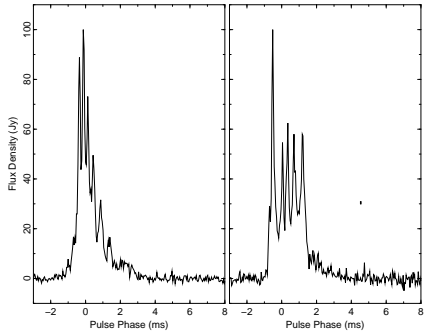,angle=-90,width=6cm}
\end{figure}

\begin{figure}
\caption{The histogram of mean flux densities observed  for 20085 pulses 
at 1413 MHz. The flux distribution is similar to other pulsars in
that very few strong pulses are seen.}
\psfig{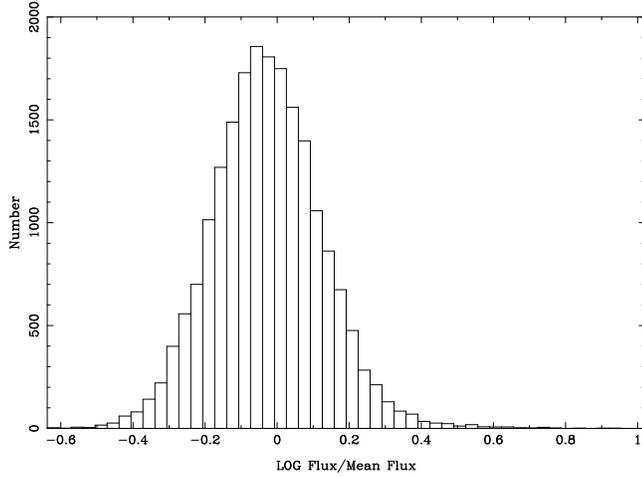}
\end{figure}

\begin{figure}
\caption{The $R$ parameter (see text) as function of pulse longitude
at 1413 MHz. Zero phase refers to the peak of the integrated profile.
The horizontal solid line represents the expected value if the intensity
were normally distributed.}
\psfig{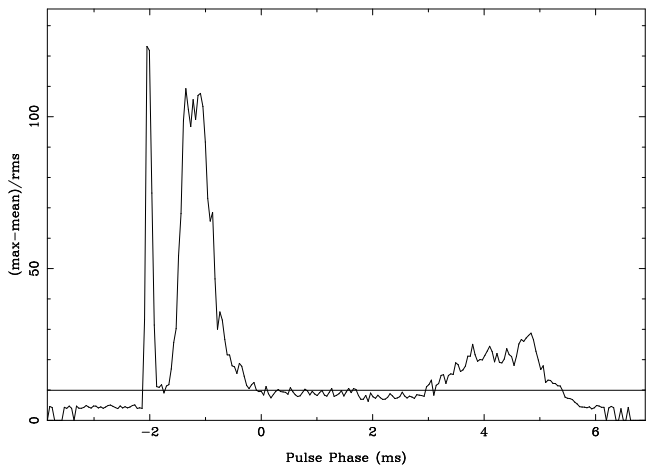}
\end{figure}

\clearpage
\newpage
\begin{figure}
\caption{The integrated profile from 21 pulses at 1413 MHz which
exhibit an additional `bump' feature at the trailing part of the profile.
The bump is orthogonally polarized compared to the main profile.
The top panel shows the swing of position angle as a function of pulse
phase. The bottom panel shows the total intensity (thick line), linear
polarization (dashed line) and circular polarization (dotted line).
The smooth thick line is the integrated profile from 20000 pulses using
the same flux density scale.}
\psfig{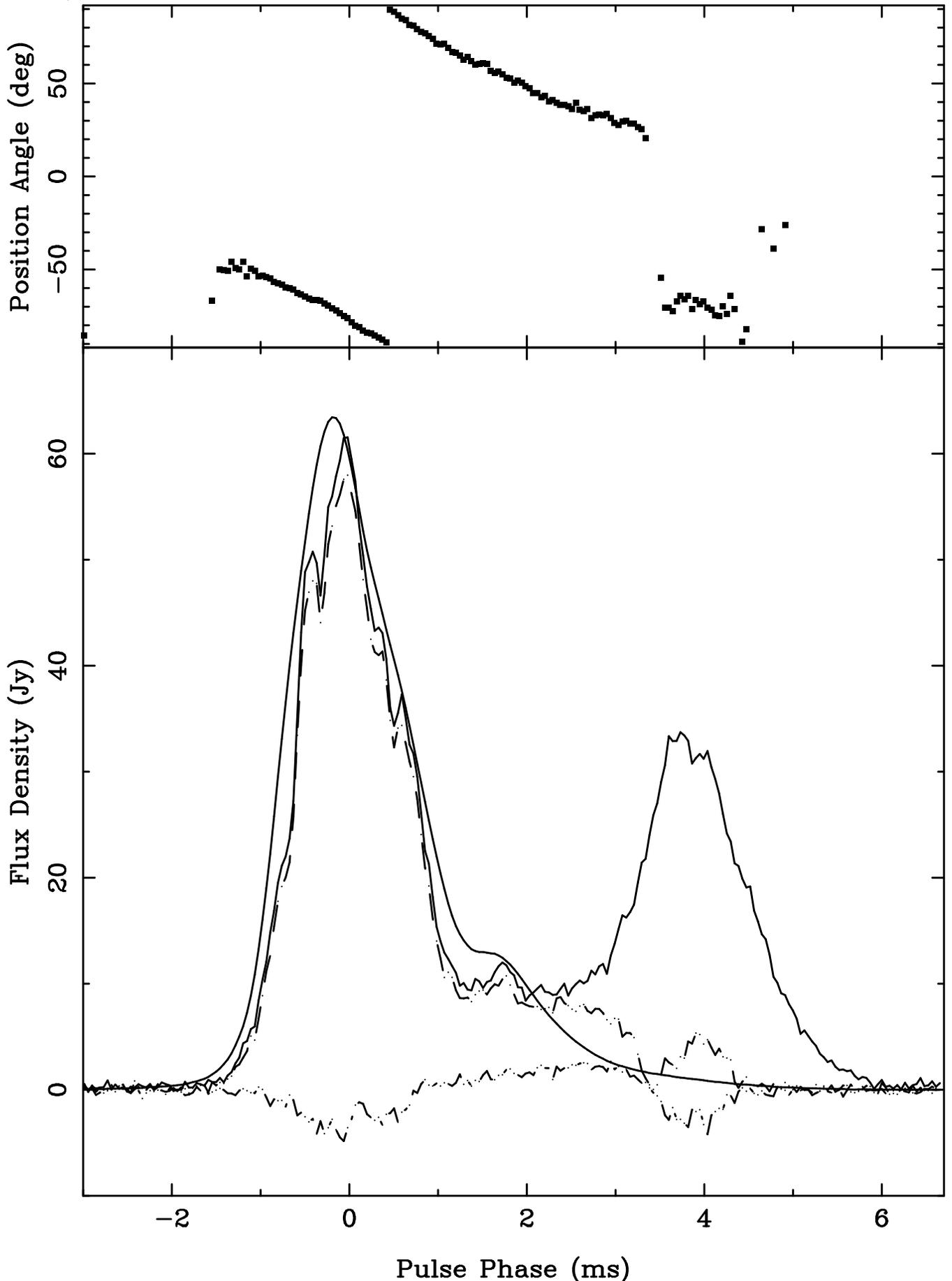}
\end{figure}

\clearpage
\newpage
\begin{figure}
\caption{A superposition of 14 giant micro-pulses at 1413 MHz occurring 
at early pulse phases. Only the total intensity is shown.
In comparison, the integrated profile has a peak flux of only 60 Jy.}
\psfig{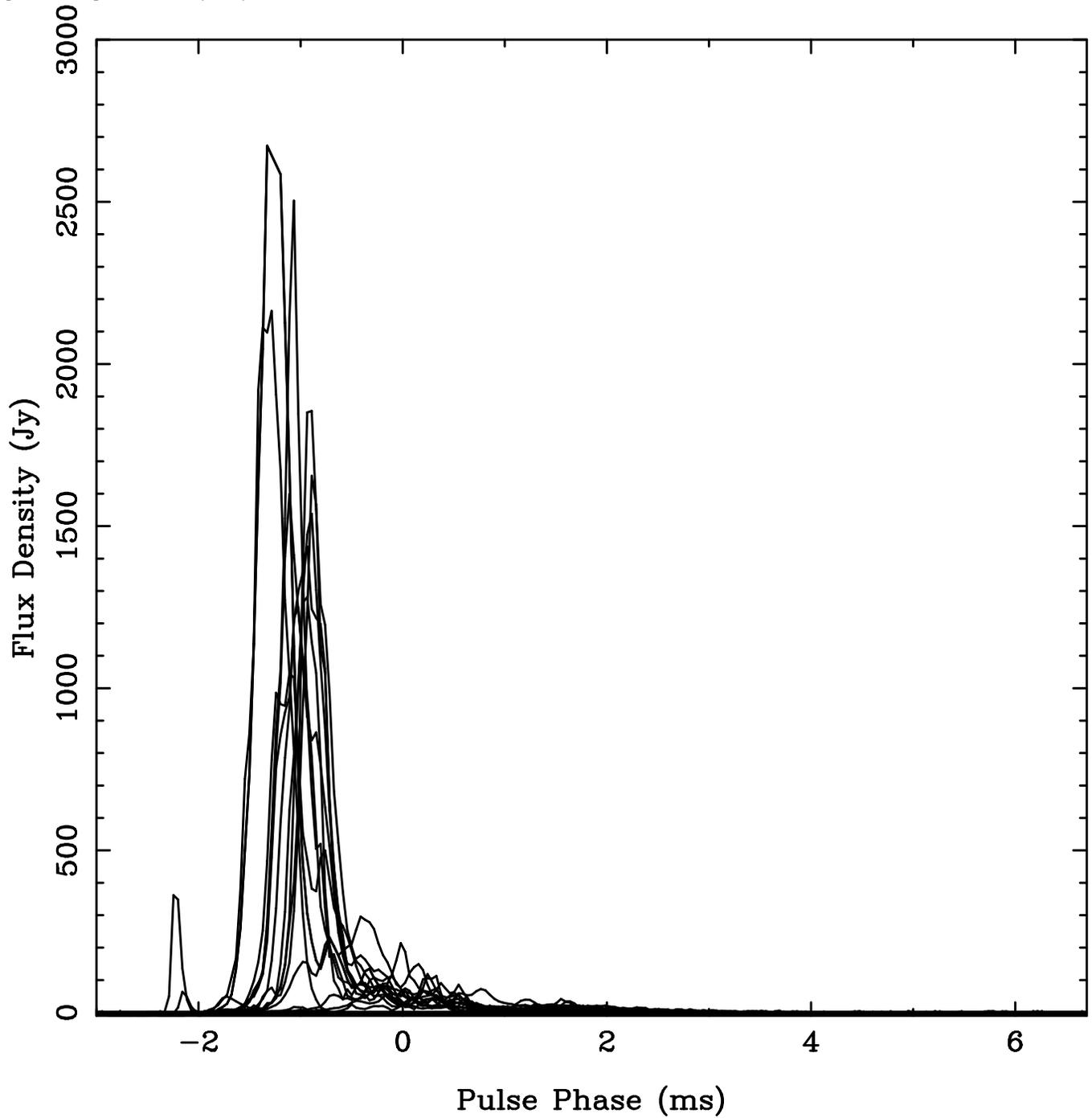}
\end{figure}

\end{document}